\documentclass[notitlepage,a4,10.5pt]{article}

\usepackage{graphicx}

\usepackage{amsmath}%
\usepackage{amsfonts}%
\usepackage{amssymb}%
\usepackage{mathrsfs}
\usepackage{amsthm}
\usepackage{float}

\usepackage{authblk}

\usepackage[left=2.5cm,right=2.5cm,top=3cm,bottom=3cm]{geometry} 

\usepackage{xcolor}

\usepackage{CJKutf8}

\usepackage{caption}    
\usepackage{subcaption}

\newcommand{\cp}{\times}

\newcommand{\bol}{\boldsymbol}

\newcommand{\abs}[1]{\left\lvert{#1}\right\rvert}

\newcommand{\lr}[1]{\left({#1}\right)}
\newcommand{\lrs}[1]{\left[{#1}\right]}
\newcommand{\lrc}[1]{\left\{{#1}\right\}}

\newcommand{\p}{\partial}

\newcommand{\ti}[1]{\textit{#1}}
\newcommand{\tb}[1]{\textbf{#1}}

\begin{document}

\title{
Numerical simulation of two-dimensional incompressible Navier-Stokes turbulence by  Clebsch potentials}
\author[1]{S. Murai}
\affil[1]{Graduate School of Frontier Sciences, \protect\\ The University of Tokyo, Kashiwa, Chiba 277-8561, Japan \protect\\ Email: sato\_naoki@edu.k.u-tokyo.ac.jp}
\author[1]{N. Sato}
\author[2]{Z. Yoshida}
\affil[2]{National Intitute for Fusion Science, \protect\\ Toki, Gifu 509‑5292, Japan\\ Email: yoshida.zensho@nifs.ac.jp}
\date{\today}
\setcounter{Maxaffil}{0}
\renewcommand\Affilfont{\itshape\small}

    \maketitle
    \begin{abstract}
 The Clebsch representation of a velocity field represents an effective tool for the analysis of physical properties of fluid flows. Indeed, a suitable choice of Clebsch potentials can be used to extract structural features that would otherwise be hidden within the complexity of fluid patterns and their evolution. In this work, we report the solution of the two-dimensional incompressible Navier-Stokes equations via Clebsch potentials. The results are in agreement with the solution of the vorticity equation for the stream function. Furthermore, we numerically demonstrate that the Shannon information entropy associated with each Clebsch potential is a growing function of time, and that it evolves at a slower rate than the rate of change in energy and enstrophy, as predicted by theory. These results pave the way for an alternative approach in the numerical study of fluid flows.
    \end{abstract}
\begin{CJK}{UTF8}{min}

\section{Introduction}

A Clebsch representation uses a suitable set of scalar functions (Clebsch potentials) to represent a given vector field. 
For example, the velocity field $\bol{v}(\bol{x},t)$ of a fluid in three-dimensional Euclidean space $\mathbb{R}^3$ can be represented as
\begin{equation}
    \bol{v} = \nabla \varphi + \sum_{i=1}^N p_i \nabla q_i,
\end{equation}
where $\varphi(\bol{x},t)$, $p_i(\bol{x},t)$, and $q_i(\bol{x},t)$, $i=1,2,...,N$, are the Clebsch potentials, $\bol{x}=\lr{x,y,z}$ and $t$ denote Cartesian coordinates and time respectively, $\nabla$ the gradient operator in $\mathbb{R}^3$, and 
the natural number $N$ expresses the number of parameters $2N+1$ used by the representation. 
The minimum number of Clebsch parameters needed to represent an arbitrary vector field depends on the dimension of the spatial domain under consideration. In particular, considering a bounded domain in Euclidean space, a two-dimensional vector field can be fully represented if $N\geq 1$, while a three-dimensional vector field requires $N\geq 2$ \cite{Yoshida09}.

As it will be clear later, a suitable choice of Clebsch parameters
enables the description of a given mechanical system through 
a set of variables that better portrays the geometrical, topological, and dynamical features of the involved vector fields, and the reduction of the governing equations to a simpler mathematical form.  
In addition to such practical advantages, Clebsch parameters also carry a fundamental theoretical significance. Indeed, 
the pairs $\lr{p_i,q_i}$, $i=1,...,N$, represent the state variables that assign the canonical Hamiltonian structure of ideal fluids \cite{YosEpi2D,Morrison98,Morrison82}: 
in these canonical variables the Euler equations are given by the action of the symplectic matrix on the gradient of the fluid Hamiltonian as a set of transport equations for the Clebsch parameters, which are advected materially by the velocity field. The canonical form is also suitable for numerical implementations that exploit the symplectic structure of the dynamical system \cite{Morrison17}.

The study of the Euler equations by means of Clebsch potentials can 
be extended to the incompressible Navier-Stokes equations \cite{Sato,Scholle16,Cartes}.
For a two-dimensional flow $\bol{v}\cdot\nabla z=0$, the velocity field $\bol{v}$ can be completely represented as 
$\bol{v} = \nabla \varphi + p \nabla q$ where the scalars $\varphi\lr{x,y,t}$, $p\lr{x,y,t}$ and $q\lr{x,y,t}$ are the Clebsch parameters. 
Then, the two-dimensional incompressible Navier-Stokes equations in a bounded region  $\Omega\subset\mathbb{R}^2$ turn into a pair of advection-diffusion equations for $p$ and $q$ and a Poisson equation for $\varphi$ (see  \cite{Sato}),
\begin{subequations}
\begin{align}
    \frac{\partial p}{\partial t}
    &= -\bol{v} \cdot \nabla p + \nu\frac{\nabla\omega\cdot\nabla p}{\omega}, \label{eq1a}\\
    \frac{\partial q}{\partial t}
    &= -\bol{v} \cdot \nabla q + \nu\frac{\nabla\omega\cdot\nabla q}{\omega} ,\label{eq1b}\\
    \Delta \varphi &= -\nabla p \cdot \nabla q -p\Delta q,\label{eq1c}
\end{align}\label{eq1}
\end{subequations}
while the mechanical pressure is given by $P=-\p_t\varphi-\bol{v}^2/2-p\p_tq$. 
In the equations above, $\nu$ is the kinematic viscosity, 
$\omega = \partial_x v_y - \partial_y v_x$ the vorticity, $\Delta=\p_x^2+\p_y^2+\p_z^2$ the Laplacian operator ($\p_z$ evaluates to zero in this setting), and the notation $\p_t=\p/\p t$ has been used for partial derivatives.
One sees that equations \eqref{eq1a} and \eqref{eq1b} consist of two elements, a material derivative and a diffusion term, while equation \eqref{eq1c}, which expresses the condition $\nabla\cdot\bol{v}=0$, determines $\varphi$ in terms of $p$ and $q$.    
We also remark that system \eqref{eq1} does not have a Hamiltonian structure due to the presence of viscosity. 

In this paper, we are concerned with the numerical solution of system \eqref{eq1}. 
Our motivation is twofold. On one hand, we wish to demonstrate the validity of the Clebsch representation as a numerical tool for the study of fluid flows, which are not necessarily limited to the Navier-Stokes and related systems. 
On the other hand, there are physical observables, which will be described in details later, that are not directly accessible from the usual Navier-Stokes equations or the corresponding vorticity equation for the stream function (on this point, see 
\cite{Sato}). Hence, we aim at calculating the time evolution of these quantities to
characterize the development of two-dimensional turbulence, which is a phenomenon encountered in several physical contexts such as geophysical fluid dynamics and plasma physics. 
In this regard, we recall that the two-dimensional incompressible Navier-Stokes equations are 
mathematically related to the
beta-plane model \cite{Charney,Rhines} describing surface flows over rotating planets 
under the effect of the Coriolis force and gravity, 
and the Hasegawa-Mima equation for the electrostatic potential in a magnetized plasma \cite{Horton,HM, Diamond}.
Indeed, these equations share the same Hamiltonian structure in the inviscid limit, 
and 
possess two inviscid invariants, fluid energy and enstrophy (or a generalized counterpart for beta-plane model and Hasegawa-Mima equation) \cite{Weinstein}.
The existence of these two invariants makes two-dimensional flows different from three-dimensional flows, because the latter do not preserve enstrophy due to vortex stretching.  
In particular, conservation of enstrophy affects the behavior of two-dimensional turbulent cascades: 
the turbulent energy spectrum  propagates from large to small wavenumbers, while enstrophy spreads from small to large wavenumbers \cite{Kolmogorov1941,Kraichnan,Batchelor, Kraichnan2}.
Such inverse turbulent cascades of energy are often 
discussed in relation 
with the formation of large scale structures and zonal flows in geophysical fluid dynamics and plasma physics  \cite{HM5}. 


As mentioned above, the Clebsch representation 
introduces a new set of physical observables with peculiar properties in addition to fluid energy and enstrophy. 
For example, by observing that equations \eqref{eq1a} and \eqref{eq1b} are formally analogous to transport equations for the `distribution functions' $p$ and $q$,  
it can be shown \cite{Sato} that under suitable boundary conditions 
the corresponding Shannon entropy measures \cite{Shannon,Jaynes} are growing functions of time, i.e.   
\begin{equation}
    \mathrm{H}[p] = -\int_\Omega p \log p \ dV, \quad \frac{d\mathrm{H}[p]}{dt} =\nu\int_{\Omega}p\abs{\nabla\log p}^2\,dV\ge 0,
\end{equation}
and similarly for $H[q]$.
The Shannon entropy measures $H[p]$ and $H[q]$
are not generally identical to thermodynamic entropy. 
However, the basic mechanism (viscous dissipation) leading to their growth is the same.
Hence, quantities like $H[p]$ and $H[q]$ can be regarded as measures of the geometrical/topological complexity of the flow $\bol{v}$, which is progressively degraded by viscosity.



In a turbulent fluid with dissipation, one expects the decay rate of functionals (energy, enstrophy, etc.) to be roughly measured by the order of the differential operators within them \cite{YosMah}, because higher order derivatives usually imply stronger gradients, which dictate the strength of diffusion. 
A classical example is the Taylor relaxation \cite{Taylor1, Taylor2} of a magnetized plasma, where magnetic energy decays faster than magnetic helicity, 
leading to the formation of self-organized plasma configurations (Beltrami equilibria  
\cite{Dombre86}).
Likewise, one may conjecture that the Shannon entropies $H[p]$ and $H[q]$ 
decay at a slower rate than energy or enstrophy. 
The rate of change of $H[p]$ and $H[q]$ as compared to 
the evolution of energy and enstrophy will therefore be one of the problem investigated in this study. 



The present paper is organized as follows.
In Sec. II, we review the systems of equations used in the subsequent numerical simulation. 
In Sec. III, we describe the detailed setting for the numerical simulation.
In sec. IV we report the results of the numerical simulation. 
In particular, we verify the consistency of the numerical solution of the 
Navier-Stokes system via Clebsch potentials with the standard vorticity equation for the stream function, show that the functionals $H[p]$ and $H[q]$ 
are growing functions of time, and compare their rate of change with that of energy and enstrophy.
Concluding remarks are given in Sec. V.

\section{2D Incompressible Navier-Stokes Equations}

\subsection{Evolution equations for Clebsch potentials and stream function}

In this study, we numerically simulated two-dimensional incompressible viscous flow by integrating the equations of motion for the Clebsch potentials, and then compared the results with the solution of the two-dimensional vorticity equation for the steam function.

The governing equations for the Clebsch potentials can be obtained
by substituting the following representation of the two-dimensional velocity field,
\begin{equation}
    \bol{v} = \nabla \varphi + p \nabla q,
\end{equation}
into the three-dimensional incompressible Navier-Stokes equations under the assumption of constant fluid density and pressure. 
Note that here $\varphi\lr{x,y,t}$, $p\lr{x,y,t}$ and $q\lr{x,y,t}$ are independent of $z$.  
The result is the system of partial differential equations 
\begin{subequations}
\begin{align}
    \frac{\partial p}{\partial t} &= -\bol{v} \cdot \nabla p 
        + \nu \frac{\nabla \omega \cdot \nabla p}{\omega},\label{pt} \\
    \frac{\partial q}{\partial t} &= -\bol{v} \cdot \nabla q
        + \nu \frac{\nabla \omega \cdot \nabla q}{\omega},\label{qt} \\
    \Delta \varphi &= - p \Delta q - \nabla p \cdot \nabla q,\label{divv}
\end{align}\label{Cl}
\end{subequations}
where $\omega=\nabla\times\bol{v}\cdot\nabla z=\nabla p\cp\nabla q\cdot\nabla z$ is the vorticity of the two-dimensional flow and $\nu$ denotes the kinematic viscosity.
Observe that in this context the continuity equation for an incompressible flow
is expressed by the divergence-free condition of the velocity field $\nabla \cdot \bol{v} = 0$, which 
corresponds to the Poisson equation \eqref{divv} for the Clebsch potential  $\varphi$. 
This equation is solved once the Clebsch parameters $p,q$ are determined from equations \eqref{pt} and \eqref{qt}. 

On the other hand, the vorticity equation for the stream function $\psi\lr{x,y,t}$ follows by substituting he velocity field $\bol{v}=\nabla\psi\cp\nabla z$ into the three-dimensional incompressible Navier-Stokes equation under the same assumptions on density and pressure. 
We have,
\begin{subequations}
\begin{align}
    \frac{\partial \omega}{\partial t} &=
    - \frac{\partial \psi}{\partial y}\frac{\partial \omega}{\partial x}
    + \frac{\partial \psi}{\partial x}\frac{\partial \omega}{\partial y}
    + \nu \Delta \omega,\\
    \omega &=-\Delta\psi
\end{align}\label{vor}
\end{subequations}
Of course, if equations \eqref{Cl} and \eqref{vor} 
are solved under the same initial and boundary conditions for the
velocity field, they must produce the same solution $\bol{v}\lr{x,y,t}$ 
for all $t> 0$. 

\subsection{The Shannon entropy of Clebsch potentials}
We now introduce certain functionals of the Clebsch potentials 
whose rate of change has a semi-definite sign under suitable boundary conditions. In particular, we are concerned with the Shannon entropy measures of the Clebsch potentials $p$ and $q$ defined by
\begin{equation}
\mathrm{H}\lrs{p}=-\int_{D}p\log p\,dV,~~~~\mathrm{H}\lrs{q}=-\int_{D}q\log q\,dV,\label{hphq}
\end{equation}
where $dV$ denotes the volume element in $\mathbb{R}^3$ and 
$D$ some bounded region in $\mathbb{R}^3$.
The time derivative of these functionals for a three-dimensional incompressible Navier-Stokes flow can be evaluated explicitly \cite{Sato}.
For example, 
\begin{equation}
    \frac{d\mathrm{H}[q]}{dt}
    = -\nu \int_{\partial D} \lrs{\nabla (q \log q) + \mathbf{A}^q \times \nabla (q \log q)} \cdot \mathbf{n}\, dS
    + \nu \int_D q |\nabla \log q |^2 dV,\label{dHqdt}
\end{equation}
where $\bol{n}$ stands for the unit outward normal to the bounding surface $\p D$, and $dS$ the surface element on $\p D$. A similar expression holds for $\mathrm{H}[p]$.
Here, $\mathbf{A}^q$ is a vector field such that 
\begin{equation}
    \frac{\partial q}{\partial t} = - \nabla \cdot (\mathbf{V}^q q),~~~~\mathbf{V}^q = \bol{v} - \nu (\nabla \log q + \mathbf{A}^q \times \nabla \log q).\label{Vq}
\end{equation}
In equation \eqref{dHqdt} one can let the boundary term vanish by setting appropriate boundary conditions on the Clebsch potential $q$, resulting in $d\mathrm{H}\lrs{q}/dt\geq 0$ whenever $q\geq 0$ (this condition can always be satisfied by shifting the Clebsch parameters by a constant; we will return to this point later on).   
Hence, the functionals $\mathrm{H}\lrs{p}$ and $\mathrm{H}\lrs{q}$
behave as entropy measures for the effective distribution functions $p$ and $q$, and therefore can be interpreted as information measures related to the geometrical and topological complexity of fluid flow. 

Let us examine how equation \eqref{dHqdt} is modified in the case of a two-dimensional flow. To this end, extend the two-dimensional bounded domain $\Omega$ in  three-dimensions as $D=\Omega\times[0,h]$, where $h$ is any real positive number, and assume that all physical quantities are independent of $z\in [0,h]$.  
Defining the Shannon entropy measures
\begin{equation}
\mathrm{H}[p]=-\int_\Omega p\log p\,dS,~~~~\mathrm{H}[q]=-\int_{\Omega}q\log q\,dS,\label{H}
\end{equation}
we have 
\begin{equation}
\begin{split}
    \frac{d\mathrm{H}[q]}{dt}
    &= -\frac{d}{dt}\int_\Omega q \log q\, dS \\
    &= -\frac{1}{h}\frac{d}{dt}\int_D q \log q\, dV \\
    &= -\frac{\nu}{h}  \int_{\partial D} \lrs{\nabla (q \log q) + \mathbf{A}^q \times \nabla (q \log q)} \cdot \mathbf{n}\, dS
    + \frac{\nu}{h}  \int_D q |\nabla \log q |^2\, dV \\
    &= -\nu\int_{\partial\Omega} \nabla \lr{q\log q} \cdot \mathbf{n}'\, dl
    + \nu\int_\Omega  \lr{1+\log q}\lr{ \frac{\nabla \omega \cdot \nabla q}{\omega} - \Delta q }\, dS
    + {\nu}\int_\Omega q |\nabla \log q |^2\, dS,
\end{split}\label{eq11}
\end{equation}
with a similar expression for $d\mathrm{H}[p]/dt$. 
Here $dl$ represents the line element on $\p\Omega$, $\bol{n}'$ the unit outward normal to $\p\Omega$, and we used the fact that in the present setting $\bol{A}^q={z\lrs{q,p}/\omega}$ with $\lrs{q,p}=\lr{\nabla q\cdot\nabla}\nabla p-\lr{\nabla p\cdot\nabla}\nabla q$ so that by standard vector identities 
\begin{equation}
\int_{\p D}\bol{A}^q\cp\nabla\lr{q\log q}\cdot{ \bol{{\rm n}}}\,dS=
\int_D\lr{1+\log q}\nabla q\cdot\nabla z\cp\lr{\frac{\lrs{q,p}}{\omega}}\,dV=
h\int_{\Omega}\lr{1+\log q}\lr{\frac{\nabla\omega\cdot\nabla q}{\omega}-\Delta q}\,dS\label{eq12}
\end{equation}
Observe that in the last line of equation \eqref{eq11}, 
the first two terms arise from the boundary integral in equation \eqref{dHqdt}, which physically represents the outflow of the entropy density $-q\log q$ from the domain $D$. In particular, observe that the second term  
comes from the top and bottom boundary surfaces $z=0$ and $z=h$ of the domain $D$ because $\bol{A}^q\cp\nabla \lr{q\log q}$ is aligned with $\nabla z$.  
In order to observe entropy growth $d\mathrm{H}\lrs{q}/dt\geq 0$ we stress again that
either boundary conditions are chosen so that the outflow of $-q\log q$ vanishes on the boundary $\p D$, 
or this outflow is subtracted from the total rate of change of $\mathrm{H}$ 
to isolate the entropy production term $\nu\int_{\Omega}q\abs{\nabla\log q}^2\, dS$.

\section{Numerical Method}
In this section we describe the 
numerical setting used to solve the evolution equations for the Clebsch potentials \eqref{Cl} and the vorticity equation \eqref{vor}.


\subsection{Normalization}
In order to simulate system \eqref{Cl}, 
it is convenient to normalize the equations as follows:
\begin{subequations}
\begin{align}
    \frac{\partial p^*}{\partial t^*}
    &= -\bol{v^*} \cdot \nabla^* p^* + \frac{1}{\mathrm{Re}}\frac{\nabla^*\omega^*\cdot\nabla^* p^*}{\omega^*},\\
    \frac{\partial q^*}{\partial t^*}
    &= -\bol{v^*} \cdot \nabla^* q^* + \frac{1}{\mathrm{Re}}\frac{\nabla^*\omega^*\cdot\nabla^* q^*}{\omega^*},\\
    \Delta^* \varphi^* &= -\nabla^* p^* \cdot \nabla^* q^* -p^*\Delta q^*,\label{phiast}
\end{align}\label{pqast}
\end{subequations}
where $\mathrm{Re}=uL/\nu $ is Reynolds number with $u$ and $L$ characteristic velocity and length, and the asterisk represents the normalized variables or differential operators.
Likewise, the vorticity equation is normalized as 
\begin{subequations}
\begin{align}
    \frac{\partial \omega^*}{\partial t^*} &=
    - \frac{\partial \psi^*}{\partial y^*}\frac{\partial \omega^*}{\partial x^*}
    + \frac{\partial \psi^*}{\partial x^*}\frac{\partial \omega^*}{\partial y^*}
    + \frac{1}{\mathrm{Re}} \Delta^* \omega^*, \\
    \omega^* &= -\Delta^* \psi^*.
\end{align}\label{vorast}
\end{subequations}
In the following, we shall omit $\ast$ to simplify the notation.

\subsection{Boundary conditions, Initial conditions, and numerical scheme}
Equations \eqref{pqast} and \eqref{vorast} are solved in the unit square $\Omega=[0,1]^2$. Each side of the square is divided into $N=512$ grid points, the   
Reynolds number is set to $\mathrm{Re}=20$, and the time step is 
 set to $\Delta t = 1 \times 10^{-5}$ .
 
We further assume the Clebsch potentials $p,q,\varphi$ to satisfy Neumann boundary conditions: on the boundary $\partial\Omega$ the normal derivatives of $p,q,\varphi$ are all set to zero:
\begin{equation}
\nabla p\cdot{\bol{{\rm n}}'}=0,~~~~\nabla q\cdot{ \bol{{\rm n}}'}=0,~~~~\nabla\varphi\cdot{ \bol{{\rm n}}'}=0~~~~{\rm on}~~\p\Omega.\label{BC1}
\end{equation}
Imposing Neumann boundary conditions on the Clebsch parameters $q$ and $\varphi$ implies that the component of the velocity field perpendicular to the bounding surface vanishes, $\bol{v}\cdot\mathbf{n}'=\lr{\nabla\varphi+p\nabla q}\cdot\mathbf{n}'=0$ on $\p\Omega$, and the velocity field is confined in $\Omega$.  
We also remark that the boundary conditions \eqref{BC1} are different from the no slip boundary condition $\bol{v}=\bol{0}$ on $\p\Omega$ encountered in the 
standard Navier-Stokes system expressed in terms of $\bol{v}$. 
Nonetheless, they are sufficient to solve system \eqref{pqast}. 

In contrast, the boundary conditions on vorticity $\omega$ and stream function $\psi$ for the vorticity equation \eqref{vorast} are chosen to be Dirichlet boundary conditions, 
\begin{equation}
\omega=0,~~~~\psi=0~~~~{\rm on}~~\p\Omega.\label{BC2}
\end{equation}
Notice that $\psi=0$ on $\p\Omega$ implies $\bol{v}\cdot\mathbf{n}'=\nabla\psi\cp\nabla z\cdot\mathbf{n}'=0$ on $\p\Omega$ as well, because the boundary $\p\Omega$ corresponds to a level set of $\psi$. 
We also observe that the boundary conditions for the Clebsch parameters \eqref{BC1} are consistent with the boundary conditions for the vorticity equation \eqref{BC2}. Indeed, in addition to $\bol{v}\cdot\mathbf{n}'=0$
both systems satisfy $\omega=0$ on $\p\Omega$, because
$\omega = \nabla p \times \nabla q \cdot \nabla z$ in the Clebsch representation and $\nabla p$ and $\nabla q$ must be parallel on $\p\Omega$.

The initial values $p_0,q_0$ of the Clebsch parameters $p,q$ are set to
\begin{subequations}
\begin{align}
    p_0&= - \cos \pi (x-0.05\sin 4\pi x) + 2,\\
    q_0&= - \cos \pi (y-0.05\sin 4\pi y) + 2.
\end{align}\label{pq0}
\end{subequations}
The initial value $\varphi_0$ of the Clebsch parameter $\varphi$ is then evaluated by solving the Poisson equation \eqref{phiast}. 
Here, we remark that in the initial values $p_0,q_0$ the positive constant $2$ is added in order to keep $p,q$ sufficiently larger than zero during the simulation. 
Recall that one can always add a constant $c\in\mathbb{R}$ to the Clebsch potentials, since the same velocity field $\bol{v}$ can be obtained by defining $p'=p+c$, $q'=q+c$, $\varphi'=\varphi+cq$. 
This is useful to evaluate the entropy measures $\mathrm{H}\lrs{p}=-\int_{\Omega} p \log p \ dV$ and $\mathrm{H}\lrs{q}=-\int_{\Omega} q \log q \ dV$, which contain a logarithmic term. 
Notice also that the initial conditions \eqref{pq0} are consistent  with the boundary conditions \eqref{BC1}, 
and that any other compatible initial conditions could be used as well.

\begin{figure}[h]
    \hspace*{-0cm}\centering
    \includegraphics[scale=0.6]{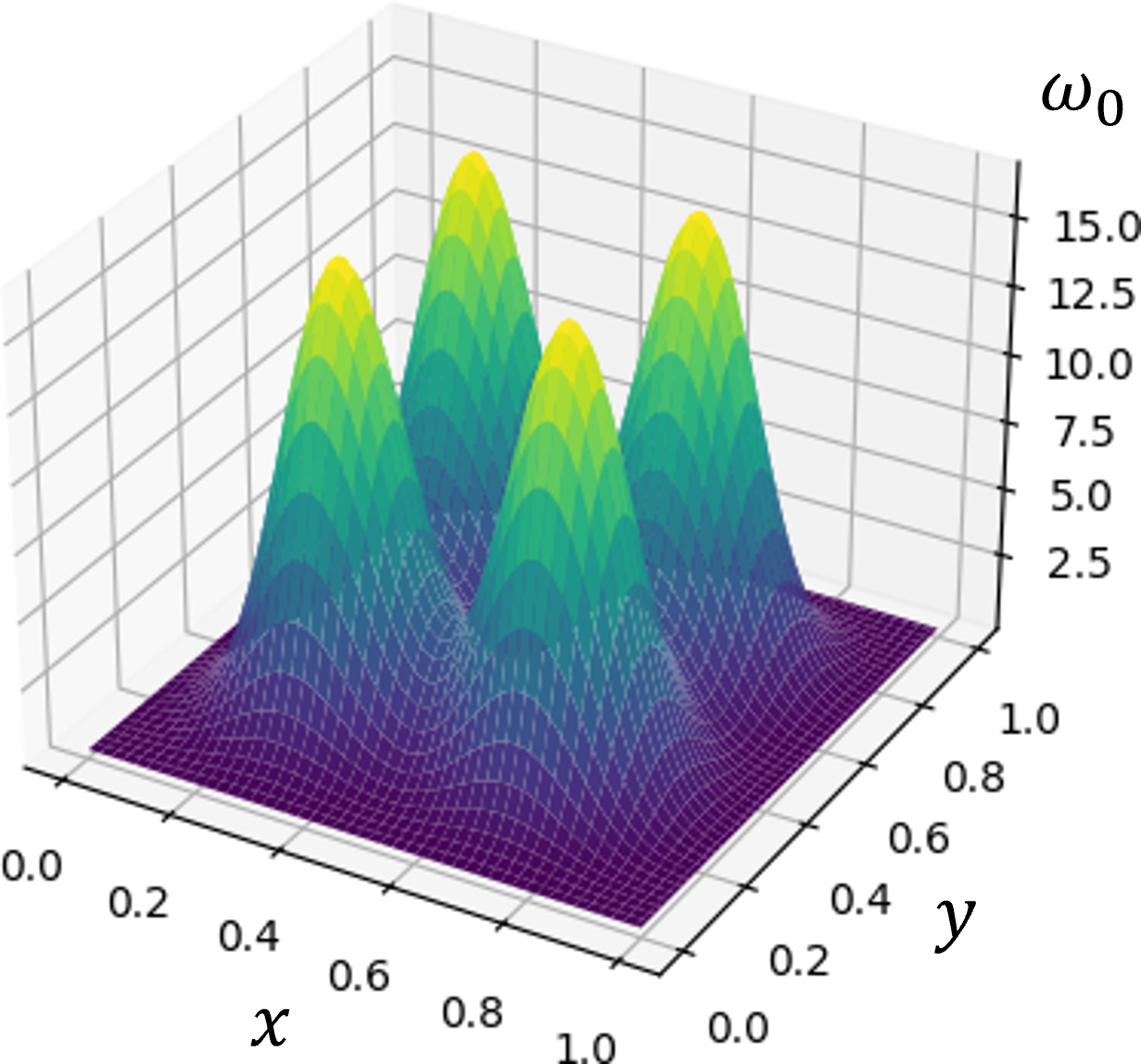}
    \caption{Initial value $\omega_0$ of the vorticity $\omega$ in the unit square $\Omega=[0,1]^2$.}
    \label{fig:omega_3D}
\end{figure}

With regard to the vorticity equation, the initial value of the vorticity $\omega$ is set to $\omega_0 = (\nabla p_0 \times \nabla q_0) \cdot \nabla z$. The initial value of $\psi$ then follows by solving the Poisson equation $\omega_0=-\Delta\psi_0$ under Dirichlet boundary conditions for $\psi_0$. 
As shown in Figure \ref{fig:omega_3D},
the initial value of the vorticity is zero at the boundary $\partial\Omega$, and is positive inside. 
The vanishing of $\omega$ on $\p\Omega$ poses a technical challenge 
with respect to the computation of the term $\nabla\omega/\omega$ occurring on the right-hand side of the first two equations in system \eqref{pqast}. 
Indeed, although the ratio $\nabla\omega/\omega$ is expected to remain finite, its numerical evaluation involves division by zero. 
To avoid this issue, we therefore shift the position of the grid points according to
\begin{subequations}
\begin{align}
    x_j &= \frac{1}{2\Delta x} + j\Delta x,~~~~j=0,1,2,\cdots,N-1, \\
    y_k &= \frac{1}{2\Delta y} + k\Delta y,~~~~k=0,1,2,\cdots,N-1,
\end{align}\label{GP}
\end{subequations}
where $\Delta x = \Delta y = 1/512$
is the interval between two adjacent grid points in both the $x$ and $y$ directions.
Observe that the grid defined by \eqref{GP} enables us to avoid 
actual computations on $\p\Omega$, and thus division by zero in the evaluation of the ratio $\nabla\omega/\omega$. 
More generally, we note that the zeros of a sufficiently regular vorticity $\omega$ could be handled in the same way by shifting 
grid points so that they never fall on the contours $\omega=0$. 

The time evolution equations \eqref{pqast} and \eqref{vorast} are solved numerically by forward differentiation in time and central differentiation in space (FTCS), 
while the Poisson equations occurring in both systems are solved numerically by discrete Fourier transform 
(more precisely, discrete sine transform for the Dirichlet problem,
and discrete cosine transform for the Neumann one).


\section{Numerical Results}
The results of the numerical simulation are reported in this section. 
\vspace{0.2 cm}

\noindent\ti{a) Time evolution of vorticity} 
\vspace{0.2 cm}

\begin{figure}[h]
    \hspace*{-0.1cm}\centering
    \includegraphics[scale=0.42]{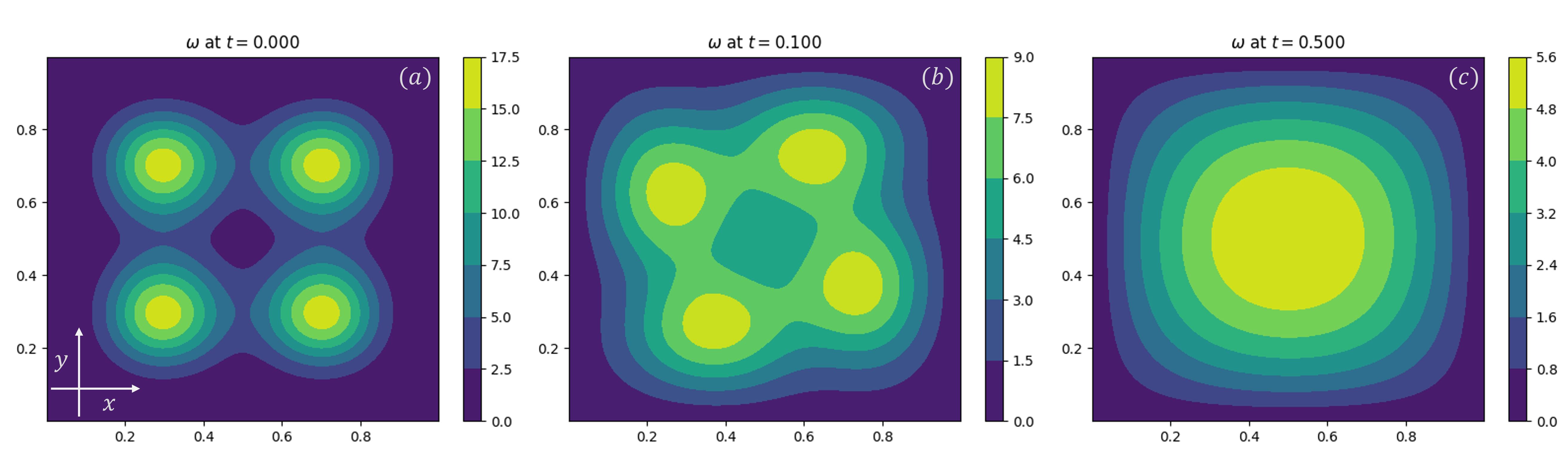}
    \caption{Distribution of vorticity $\omega$ at (a) $t=0$, (b) $t=0.1$, and (c) $t=0.5$.}
    \label{fig:omega_trio}
\end{figure}

\noindent The time evolution of the vorticity $\omega$ obtained by solving system \eqref{pqast} for the Clebsch formulation of the two-dimensional incompressible Navier-Stokes equations under the boundary conditions and the initial conditions described in the previous section is shown in Figure \ref{fig:omega_trio}. 
The evolution of $\omega$ occurs in two separate phases. 
Initially, the nonlinear term $\bol{v}\cdot\nabla\bol{v}$ (or equivalently the term $\psi_x\omega_y-\psi_y\omega_x$ in the vorticity formulation \eqref{vorast}) 
is dominant, and the four vortices contained in $\omega_0$ are advected counter-clockwise (compared figure 2(a) and figure 2(b)). 
After some time the nonlinear term 
becomes progressively smaller, until diffusion driven by the viscous term dominates. 
At this point advection fades out, and the 
vorticity is diffused uniformly (see figure 2(c)). 
In particular, observe that 
the maximum value of the vorticity $\omega$ is a decreasing function of time.

\begin{figure}[h]
    \hspace*{-0.1cm}\centering
    \includegraphics[scale=0.41]{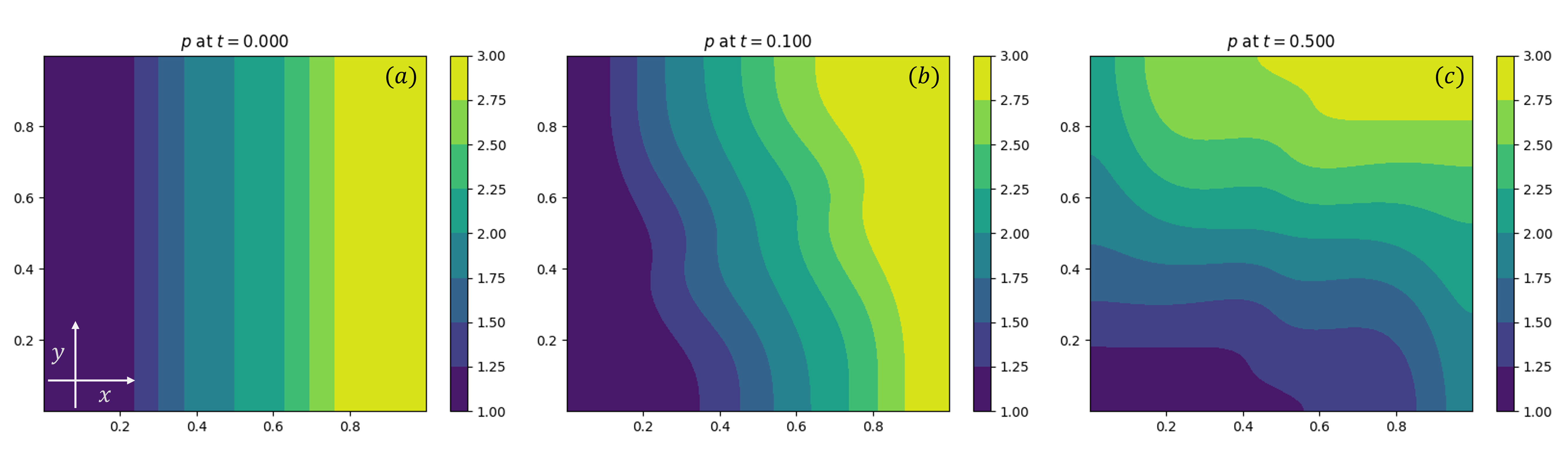}
    \caption{Distribution of the Clebsch parameter $p$ at (a) $t=0$, (b) $t=0.1$, and (c) $t=0.5$.}
    \label{fig:p_trio}
\end{figure}

\begin{figure}[h]
    \hspace*{-0.1cm}\centering
    \includegraphics[scale=0.41]{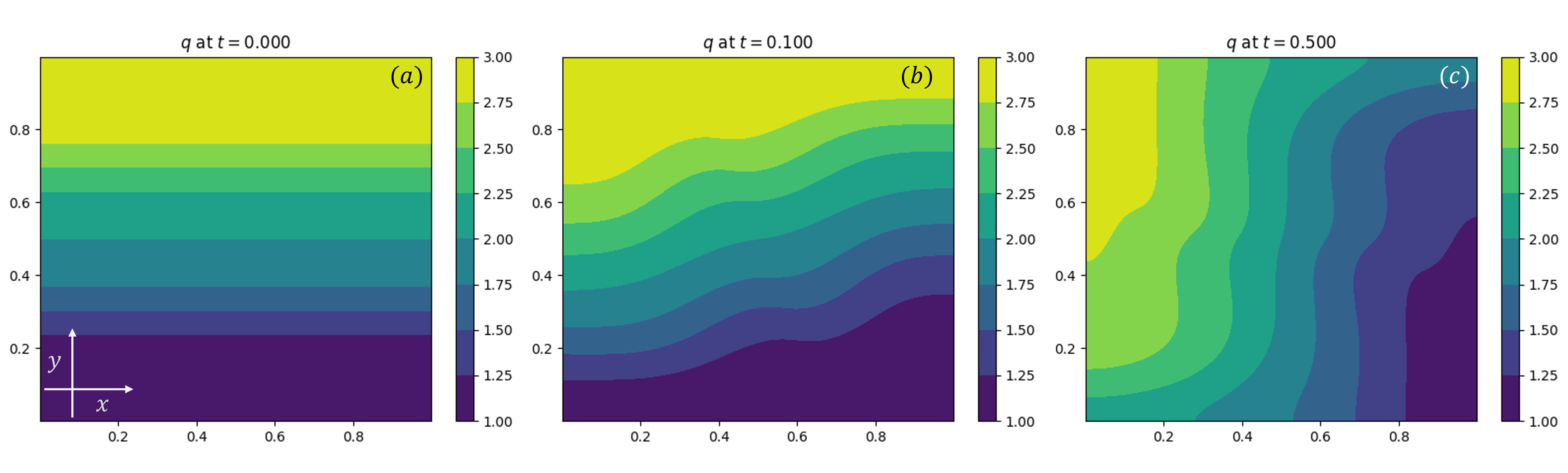}
    \caption{Distribution of the Clebsch parameter $q$ at (a) $t=0$, (b) $t=0.1$, and (c) $t=0.5$.}
    \label{fig:q_trio}
\end{figure}



\vspace{0.2 cm}
\noindent\ti{b) Time evolution of Clebsch potentials}
\vspace{0.2 cm}

\noindent 
We now focus on the time evolution of the Clebsch potentials 
obtained from system \eqref{pqast}.  
The time evolution of the potential $p$ 
is shown in figure \ref{fig:p_trio}, while that of $q$ is shown in figure \ref{fig:q_trio}. 
Both Clebsch potentials $p$ and $q$ are advected by the  velocity field in the counter-clockwise direction as in the case of the vorticity $\omega$ in figure \ref{fig:omega_trio}.
However, we observe that the advection phase of the Clebsch pontentials $p$ and $q$ persists for a longer time interval. 
We will see that this fact is related to 
the growth rate of the entropies $\mathrm{H}\lrs{p}$ and $\mathrm{H}\lrs{q}$, which grow at a slower pace
than the rate at which energy and enstrophy decay. 
We also observe that the contours of $p$ and $q$ 
exhibit a different topology compared to the profile of $\omega=\nabla p\cp\nabla q\cdot\nabla z$. This is because $p$ and $q$ are related to $\omega$ only through their  gradients $\nabla p$ and $\nabla q$.   



\vspace{0.2 cm}
\noindent \ti{c) Errors}
\vspace{0.2 cm}

\noindent To ensure that the numerical simulation of system \eqref{pqast} produces the correct solution of the two-dimensional incompressible Navier-Stokes equations, 
we evaluated the relative error $\mathrm{RE}$ of the vorticity field $\omega$ by comparing the result of the simulation of system \eqref{pqast} for the Clebsch potentials 
with the numerical solution of the standard vorticity equation \eqref{vorast}. 
%
The relative error $\mathrm{RE}$ is defined by 
\begin{equation}
    \mathrm{RE} = \sqrt{\frac{
        \sum_{j,k=1}^N \lr{\omega_\mathrm{Clebsch}^{jk}- \omega_\mathrm{Control}^{jk}}^2
    }{
       \sum_{j,k=1}^N \lr{\omega_\mathrm{Control}^{jk}}^2
    }},
\end{equation}
where $\omega_\mathrm{Clebsch}^{jk}$ is the vorticity 
at the grid point $\lr{j,k}$ 
obtained by simulating the Clebsch system \eqref{pqast},
and $\omega_\mathrm{Control}^{jk}$ is the vorticity at the grid point $\lr{j,k}$ obtained by simulating the vorticity equation \eqref{vorast}.
\begin{figure}[h!]
    \centering
    \includegraphics[scale=0.55]{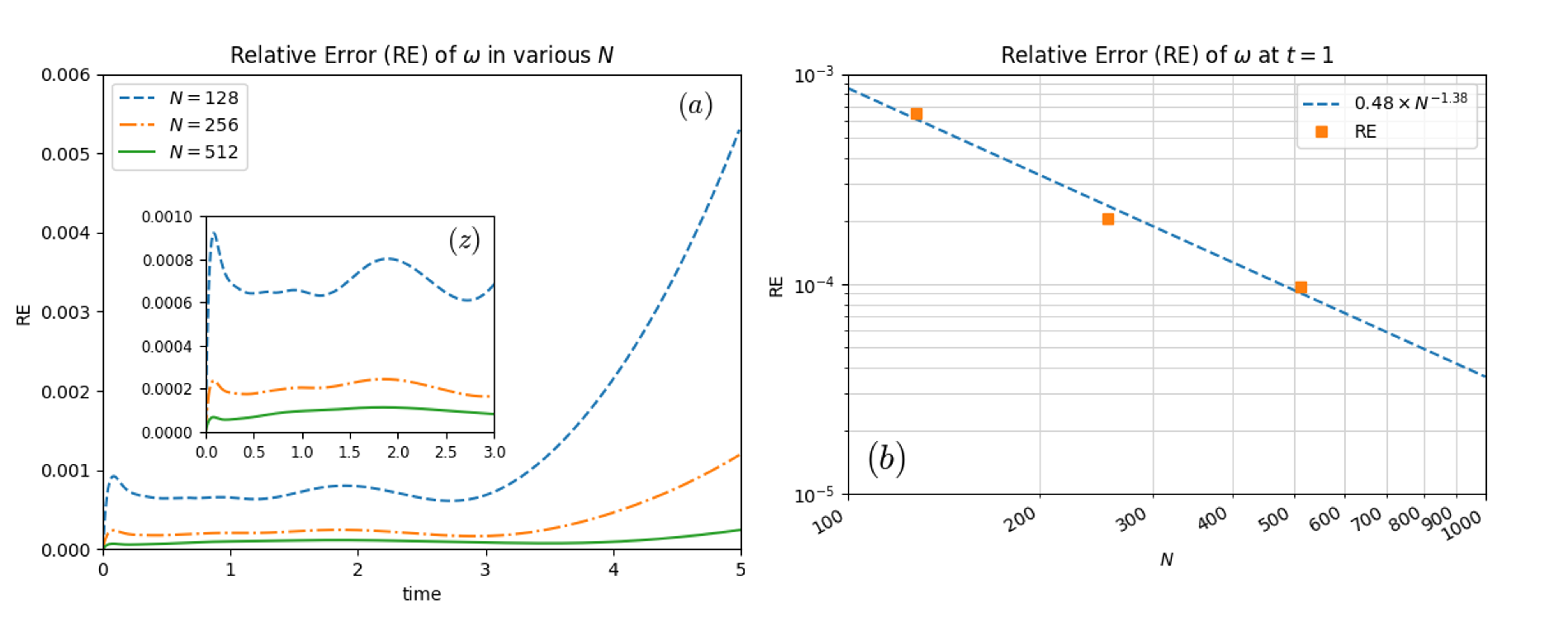}
    \caption{(a) The time evolution of the relative error $\mathrm{RE}$ of the vorticity $\omega$ 
    obtained by comparing the Clebsch model \eqref{pqast} and the standard vorticity equation \eqref{vorast}
    for various mesh numbers $N$. 
    The blue dashed line represents $N=128$, 
    the orange dashed-dotted line $N=256$,
    and the green solid line $N=512$. (b) Graph of the relative error $\mathrm{RE}$ at the instant $t=1$ as a function of $N$.}
    \label{fig:RE}
\end{figure}
Figure \ref{fig:RE}(a) shows the time evolution of the relative error $\mathrm{RE}$ for different values of the mesh number $N$. 
As $N$ becomes larger, $\mathrm{RE}$ becomes smaller. 
Therefore, we conclude that there is no systematic error arising from the simulation model \eqref{pqast} using Clebsch potentials. 
Figure \ref{fig:RE}(z) shows the change in $\mathrm{RE}$ over the time interval $0\leq t\leq 3$.  
During this interval, the value of $\mathrm{RE}$ is below  0.1\% for all the values of $N$ used in the simulation. When $t \geq 3$, $\mathrm{RE}$ becomes to increase due to the long term time integration in which errors are  accumulated continuously. 
Notice that this effect is mitigated by a larger value of $N$ as well. 
Figure 5(b) is a graph comparing the relative error $\mathrm{RE}$ at the instant $t=1$ for different values of $N$. We find $\mathrm{RE}\approx 0.5\cp N^{-1.4}$. 

\vspace{0.2 cm}
\noindent \ti{d) Time evolution of the Shannon entropy measures}
\vspace{0.2 cm}

\noindent 
We now consider the time evolution of the Shannon entropy entropy measures $\mathrm{H}[p]$ and $\mathrm{H}[q]$ defined in \eqref{H}. 
In the present setting, the rates of change in $\mathrm{H}[p]$ and $\mathrm{H}[q]$ are made of two contributions. 
The first one originates from the boundary integral in \eqref{eq11}, and it does not have a definite sign. 
Physically, it represents entropy loss at the boundary. 
The second one arises from the volume integral in \eqref{eq11} and it is always nonnegative. 
In particular, the same calculation used in equations \eqref{eq11} and \eqref{eq12} combined with the boundary conditions for the Clebsch potentials \eqref{BC1} gives 
\begin{subequations}
\begin{align}
    \frac{d\mathrm{H}[p]}{dt}
    &= \nu\int_\Omega (1 + \log p) \left( \frac{\nabla \omega \cdot \nabla p}{\omega} - \Delta p \right)\,dS
    + \nu\int_\Omega p |\nabla \log p |^2\,dS,\label{eq20a} \\
    \frac{d\mathrm{H}[q]}{dt}
    &= \nu\int_\Omega (1 + \log q) \left( \frac{\nabla \omega \cdot \nabla q}{\omega} - \Delta q \right)\,dS
    + \nu\int_\Omega q |\nabla \log q |^2\,dS.
\end{align}
\end{subequations}
Notice that the first term on the right-hand side of both equations can be written as a boundary integral on $\p D$  
(recall that $D=\Omega\cp[0,h]$), while the second term on the right-hand side is nonnegative whenever the Clebsch potentials $p$ and $q$ are nonnegative. This latter condition can always be satisfied by shifting $p$ and $q$ by a constant $c\in\mathbb{R}$ as already discussed in section 3.2, and it is fulfilled in our simulation through the constant $+2$ on the right-hand side of the initial conditions \eqref{pq0}.

\begin{figure}[h!]
    \hspace*{-0.1cm}\centering
    \includegraphics[scale=0.65]{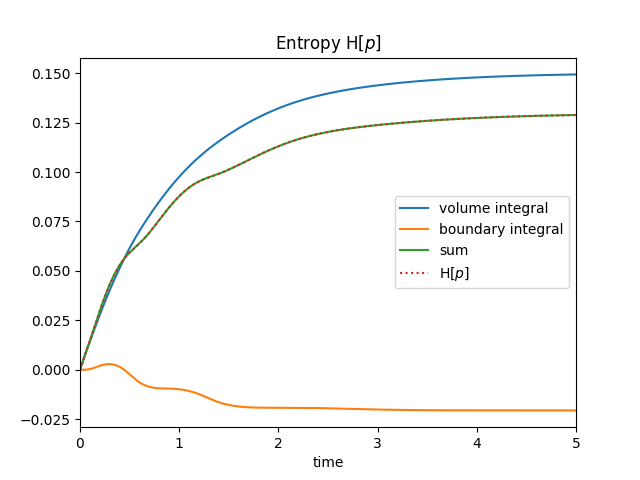}
    \caption{Time evolution of the entropy measure $\mathrm{H}[p]$. The red dotted line (which  overlaps with the solid green line) is obtained by plotting the integral $\mathrm{H}[p]=-\int_{\Omega}p\log p$ as a function of time $t$.
    The blue solid line is the accumulated value of the entropy production term (second term on the right-hand side of \eqref{eq20a}), 
    while the orange solid line corresponds to  
    the accumulated value of the boundary loss term (first term on the right-hand side of \eqref{eq20a}). 
    The Green solid line represents the summation of these two contributions, which coincides with the red dotted line.} 
    \label{fig:entropy_p}
\end{figure}

The time evolution of the Shannon entropy measure $\mathrm{H}[p]$ is shown in figure \ref{fig:entropy_p} (for the graph of $\mathrm{H}[q]$ see figure 7). 
The time evolution of the contributions due to the  boundary integral and the volume integral on the right-hand side of \eqref{eq20a} are also shown separately. 
We see that the boundary integral is not always positive, 
resulting in a net loss of entropy at the boundary $\p D$. This result is consistent with the fact that the effective velocity field advecting the Clebsch potentials $p$ and $q$ is generally different from the fluid velocity $\bol{v}$, as described by equation \eqref{Vq}.
Therefore, the tangential boundary condition $\bol{v}\cdot\bol{n}'=0$ is not enough to prevent $\mathrm{H}[p]$ and $\mathrm{H}[q]$ from being lost at the bounding surface $\p D$. 
Nonetheless, the entropy production terms $\int_{\Omega} p\abs{\nabla\log p}^2\,dS$ and
$\int_{\Omega} q\abs{\nabla\log q}^2\,dS$
are always positive during the simulation, resulting in
maximization of $\mathrm{H}[p]$ and $\mathrm{H}[q]$ as predicted by theory. 


\vspace{0.2 cm}
\noindent \ti{e) Decay rates of enstrophy, energy, and entropies}\\

\noindent 
In this paragraph we are concerned with the speed of decay of enstrophy $W$ and energy $E$ as compared with the rate of increase in the Shannon entropy measures $\mathrm{H}[p]$ and $\mathrm{H}[q]$ and the speed of decay of other functionals of the Clebsch potentials $p$ and $q$ that we will introduce shortly. 
We recall that in this context enstrophy and energy are respectively defined by the $L^2\lr{\Omega}$ norms 
\begin{equation}
W=\frac{1}{2}\int_{\Omega}\omega^2\,dS,~~~~E=\frac{1}{2}\int_{\Omega}\bol{v}^2\,dS.
\end{equation}
We also introduce the $L^2\lr{\Omega}$ norms $P$ and $Q$ of the Clebsch potentials $p$ and $q$ according to
\begin{equation}
P=\int_{\Omega}p^2\,dS,~~~~Q=\int_{\Omega}q^2\,dS. 
\end{equation}
As shown in \cite{Sato}, these norms 
are decreasing functions of time under appropriate boundary conditions. 

\begin{figure}[h!] 
    \centering
    \includegraphics[scale=0.54]{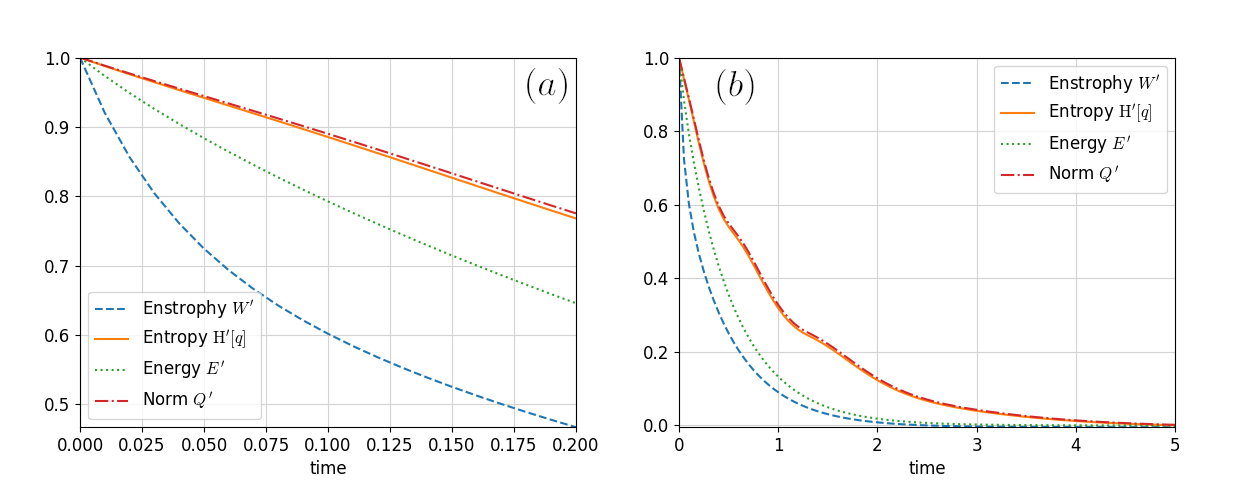}
    \caption{Decay rates of enstrophy $W$, energy $E$, negative entropy measure $-\mathrm{H}[q]$ and norm $Q$ during (a) the turbulent phase $0 \le t \le 0.2$ and (b) the entire time interval of the simulation $0 \le t \le 5$.
    The curves in the graphs  show how these observables decayed compared to their initial values   
    $W(0), -\mathrm{H}[q](0), E(0)$ and $Q(0)$ 
    through the quantities 
    $W'=(W-\beta_W)/(W(0)-\beta_W), \mathrm{H}'[q]=(\mathrm{H}[q]+\beta_q)/(\mathrm{H}[q](0)+\beta_q),
    E'=(E-\beta_E)/(E(0)-\beta_E)$, and $Q'=(Q-\beta_Q)/(Q(0)-\beta_Q)$.
    Here, $\beta_W, \beta_q, \beta_E$ and $\beta_Q$ are the constant terms in the fitting curves
    $\alpha e^{-t/\gamma}+\beta$ with $\alpha,\beta,\gamma\in\mathbb{R}$ for the decay curves of 
    enstrophy $W$, energy $E$, negative entropy measure $-\mathrm{H}[q]$ and norm $Q$.
    }
    \label{fig:diffusion}
\end{figure}

\begin{table}[h!]
        \centering
    \begin{tabular}{lcc}
        Obervable & Expression & Time constant $\gamma$ \\
        \hline
        \hline
        Enstrophy $W$ & $\frac{1}{2}\int_\Omega \omega^2 \ dV$ & $0.198$ \\
        Energy $E$ & $\frac{1}{2}\int_\Omega \bol{v}^2 \ dV$ & $0.430$ \\
        Entropy $\mathrm{H}[q]$ & $-\int_\Omega q \log q \ dV$ & $0.828$ \\
        Norm $Q$ & $\int_\Omega q^2 \ dV$ & $0.864$ \\
        \hline
    \end{tabular}
    \caption{Time constants of enstrophy $W$, energy $E$, entropy measure $\mathrm{H}[q]$ and norm $Q$ during the turbulent phase $0 \le t \le 0.2$. A smaller time constant results in faster decay of the corresponding  observable.}
    \label{table:time_constants}
\end{table}


Figure \ref{fig:diffusion} shows the time evolution of enstrophy $W$, energy $E$, Shannon entropy measure  $\mathrm{H}[q]$, and the norm $Q$ 
during the turbulent phase $0 \le t \le 0.2$ in which the nonlinear advection $\bol{v}\cdot\nabla\bol{v}$ is still dominant. 
It can be seen from this figure that the slope of the graphs are different.  
In particular, enstrophy $W$ decays at the fastest rate, followed by energy $E$, entropy measure $\mathrm{H}[q]$, and the norm $Q$.
This result is {also explained quantitatively in table \ref{table:time_constants} in terms of the time constant  $\gamma$ obtained by fitting the decay curves with the function $\alpha e^{-t/\gamma}+\beta$ with $\alpha,\beta,\gamma\in\mathbb{R}$ and is}   
consistent with the expectation that when the flow is in a turbulent state
the quantities involving higher order spatial derivatives (e.g. enstrophy) 
decrease at a faster pace than functionals containing  lower order derivatives of the dynamical variables (e.g. $\mathrm{H}[q]$). This is because higher-order derivatives imply stronger gradients, which in turn result in stronger diffusion.  

\vspace{0.2 cm}
\noindent \ti{f) Changing initial conditions}\\

\begin{figure} 
    \centering
    \includegraphics[scale=0.41]{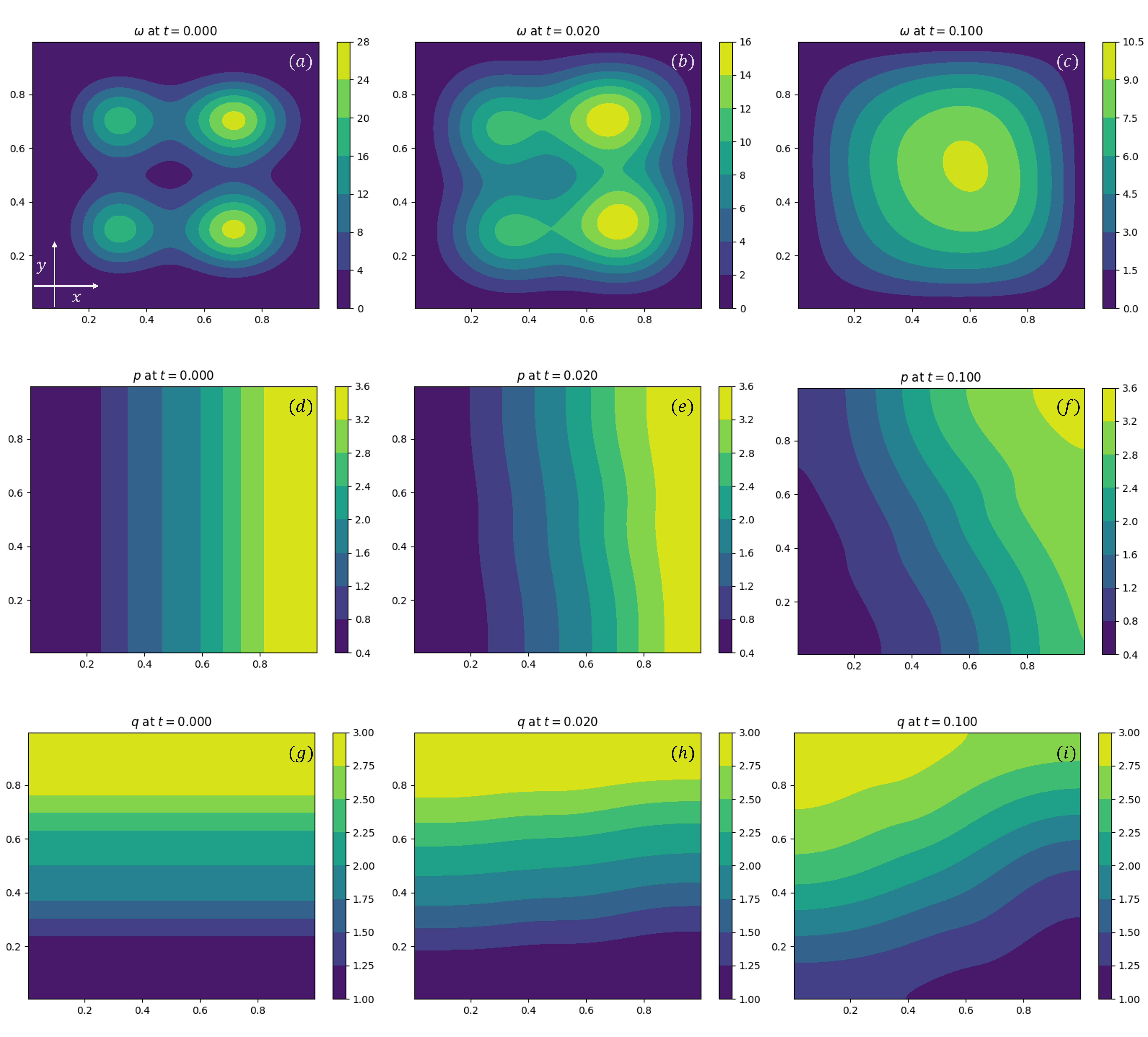}
    \caption{Distribution of vorticity $\omega$ at (a) $t=0$, (b) $t=0.02$, and (c) $t=0.1$. Distribution of the Clebsch parameter $p$ at (d) $t=0$, (e) $t=0.02$, and (f) $t=0.1$. 
    Distribution of the Clebsch parameter $q$ at (g) $t=0$, (h) $t=0.02$, and (i) $t=0.1$.
    }
    \label{fig8}
\end{figure}

\begin{figure} 
    \centering
    \includegraphics[scale=0.54]{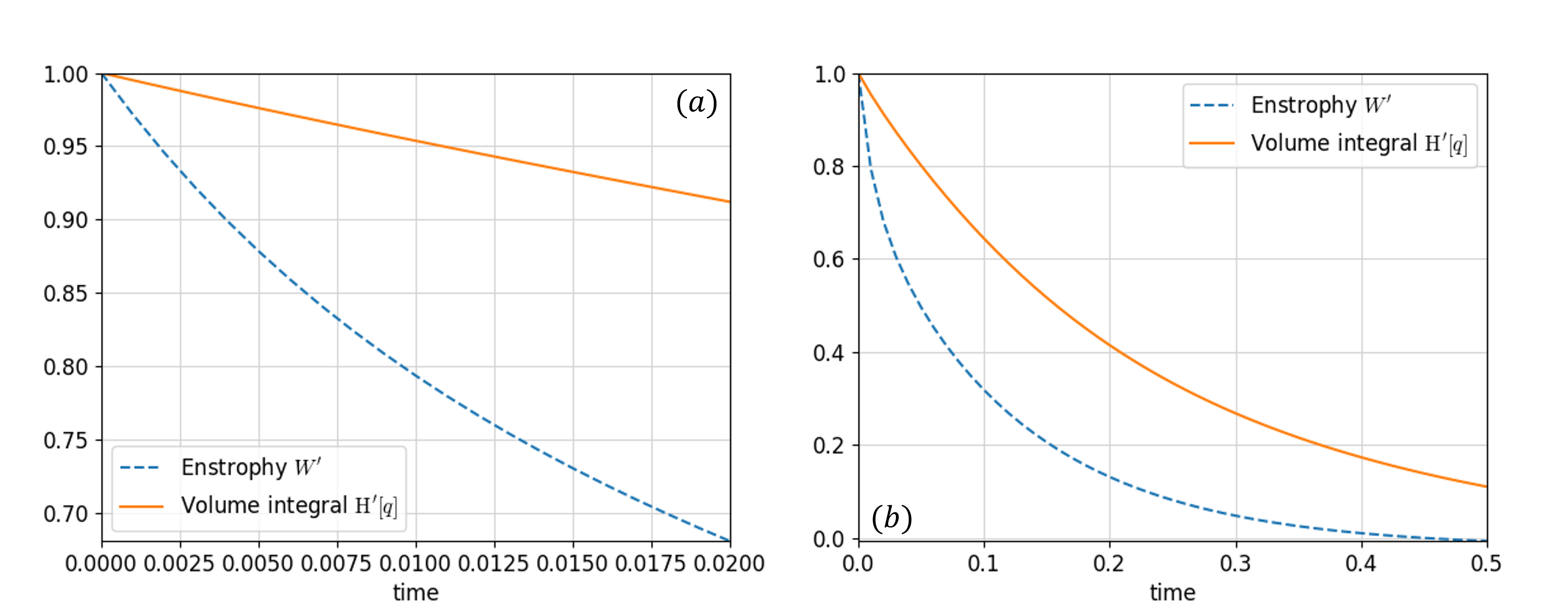}
    \caption{Decay rates of enstrophy $W$ and the volume contribution to the negative entropy measure $-\mathrm{H}_V[q]$  during (a) the interval $0 \le t \le 0.02$ and (b) the time interval $0 \le t \le 0.5$. 
    The curves in the graphs show how these observables decayed compared to their initial values   
    $W(0)$ and $-\mathrm{H}[q](0)$  
    through the quantities 
    $W'=(W-\beta_W)/(W(0)-\beta_W)$ and $\mathrm{H}'[q]=(\mathrm{H}_V[q]+\beta_q)/(\mathrm{H}[q](0)+\beta_q)$ where $\mathrm{H}_V\lrs{q}$  represents the accumulated value of the entropy production term (volume integral) in \eqref{dHqdt} at a given instant. 
    Here, $\beta_W$ and $\beta_q$ are the constant terms in the fitting curves
    $\alpha e^{-t/\gamma}+\beta$ with $\alpha,\beta,\gamma\in\mathbb{R}$ for the decay curves of 
    enstrophy $W$ and volume contribution to the negative entropy measure $-\mathrm{H}_V[q]$. 
    }
    \label{fig9}
\end{figure}

In this last paragraph
we consider how a different set of initial conditions affects the results reported above. More precisely, we modify \eqref{pq0}
as follows:
\begin{subequations}
\begin{align}
    p_0&= - \cos \pi (x-0.05\sin 4\pi x) 
    -\frac{1}{2}\cos\lrc{\pi\lrs{1-\cos\lr{\frac{\pi}{2}x}}}
    +2,\\
    q_0&= - \cos \pi (y-0.05\sin 4\pi y) + 2.
\end{align}\label{pq02}
\end{subequations}
The time evolution of vorticity $\omega$ and Clebsch parameters $p$ and $q$
is given in figure \ref{fig8}, while the corresponding 
decay rates of enstrophy $W$ and entropy measure $\mathrm{H}[q]$
can be found in figure \ref{fig9}. 
The results of the simulation are analogous to those obtained
from the original set of initial conditions \eqref{pq0}, a fact that   
shows the degree of robustness of the numerical scheme based on 
the Clebsch representation of the velocity field 
with respect to different flow configurations.


\section{Concluding Remarks}
The Clebsch representation of a vector field can be used to extract geometric, topological, and turbulent properties of fluid flows that are often inaccessible by other means. In this work, we numerically solved the Clebsch formulation of the two-dimensional incompressible Navier-Stokes equations and characterized the evolution of turbulence through the Shannon entropy measures of the Clebsch potentials in addition to energy and enstrophy. The accuracy of the numerical approach was verified by direct comparison with the solution of the standard vorticity equation for the stream function. Furthermore, we showed that the Shannon entropy measures of the Clebsch potentials increase with time, and that their rate of change is slower than the rate of decay of energy and enstrophy, as expected from theory. These findings demonstrate the usefulness of Clebsch potentials in the understanding of turbulence and pave the way to a different approach to the numerical simulation of fluid flows. In this regard, we remark that the Clebsch representation of the velocity field can be used to model fluid flows with complex geometries, such as the flow of a liquid on a curved surface (see e.g. \cite{SatoYamada}).
We therefore envisage the application of the approach developed in the present study to different problems in fluid mechanics. 


\section*{Acknowledgment}
N.S. is grateful to M. Yamada for useful discussion.

\section*{Statements and declarations}

\subsection*{Data availability}
The data that support the findings of this study are available from the corresponding author, upon reasonable request.

\subsection*{Funding}
The research of N.S. was partially supported by JSPS KAKENHI Grant No. 21K13851 and 22H04936.

\subsection*{Competing interests} 
The authors have no competing interests to declare that are relevant to the content of this article.


\end{CJK}


\end{document}